\documentstyle[multicol,eqsecnum,prd,aps,graphicx]{revtex}

\begin{document}
\draft



\title{Lepton Flavor Violating $Z$ Decays in the Zee Model}

\author{\bf Ambar Ghosal\footnote{E-mail address: 
gp1195@mail.a.u-shizuoka-ken.ac.jp}, 
Yoshio Koide\footnote{E-mail address: koide@u-shizuoka-ken.ac.jp} 
and Hideo Fusaoka\footnote{E-mail address: fusaoka@aichi-med-u.ac.jp
}$^{(a)}$}
\address{
Department of Physics, University of Shizuoka, 
52-1 Yada, Shizuoka 422-8526, Japan \\
(a) Department of Physics, Aichi Medical University,  
Nagakute, Aichi 480-1195, Japan}
\date{\today}

\maketitle
\begin{abstract}
We calculate lepton flavor violating (LFV) $Z$ decays 
$Z\rightarrow {{e_i^\pm}}e_j^\mp$ ($i, j = e, \mu, \tau$ ; $i\neq j$) in the Zee model keeping in view 
the radiative leptonic decays 
$e_i\rightarrow e_j\gamma$ ( $i = \mu$, $\tau$ ; $j =  e$, $\mu$ ; 
$i\neq j$ ), $\mu$ decay and anomalous muon magnetic moment 
($\mu$AMM). 
We investigate three different cases of Zee $f_{ij}$ coupling 
(A) $f_{e\mu}^2$ = $f_{\mu\tau}^2$ = $f_{\tau e}^2$, 
(B) $f_{e\mu}^2$ $\gg$ $f_{\tau e}^2$ $\gg$ $f_{\mu\tau}^2$, and 
(C) $f_{\mu\tau}^2$ $\gg$ $f_{e\mu}^2$ $\gg$ $f_{\tau e}^2$ subject 
to the neutrino phenomenology. Interestingly, we find that, although 
the case (C) satisfies the large excess value of $\mu$AMM, however, 
it is unable to explain the solar neutrino experimental result, 
whereas the case (B)  satisfies the bi-maximal neutrino mixing 
scenario, but confronts with the result of $\mu$AMM experiment. 
We also find that among all the three cases, only the case (C) 
gives rise to largest contribution to the ratio 
$B(Z\rightarrow e^\pm\tau^\mp)$/$B(Z\rightarrow \mu^\pm \mu^\mp)$ 
$\simeq$ 
${10}^{-8}$ which is still two order less than the accessible 
value to be probed by the future linear colliders, 
whereas for the other two cases, this ratio is 
too 
low to be observed even in the near future for all possible LFV 
$Z$ decay modes.
\end{abstract}



\begin{multicols}{2}

\narrowtext

\section{Introduction}
The high statistics results of the SuperKamiokande (SK) 
atmospheric neutrino experiment \cite{sk} and the solar neutrino 
experiment \cite{solar} have strengthen the conjecture of 
neutrino flavor oscillation from one species to another. 
The phenomena of neutrino oscillation leads to non-zero neutrino 
mass and the scale of which is $\sim$ eV predominantly set by 
the atmospheric and solar neutrino experimental results. 
Such a tiny neutrino mass could be generated by several ways, 
namely, see-saw mechanism \cite{see}, non-renormalizable 
operators \cite{non} or through the radiative ways at the one or 
two loop level. One of the most well known model of radiative 
neutrino mass generation is proposed by Zee \cite{zee} in which 
small neutrino mass is generated at the one loop level due to 
charged scalar exchange through explicit lepton number violation. 
The model has been investigated by many authors [6 - 11]. 
The model contains one extra charged singlet scalar field with non-zero 
lepton number and another doublet Higgs field in addition 
with the standard model (SM) contents.  
The scalar field content of the Zee model is not only responsible 
to generate tiny neutrino masses  but also gives rise 
to non-standard interactions due to the presence of 
charged scalar fields, e.g., one of them is the anomalous muon 
magnetic moment ($\mu$AMM). The excess value of ($\mu$AMM), 
$\delta a_\mu$ = (43$\pm$ 16)$\times$ ${10}^{-10}$ recently 
reported by E821 Collaboration \cite{e821} leads to 
the new source of interactions beyond the SM level. 
Furthermore, the Zee model leads to possible Lepton Flavor Violating 
(LFV) $Z$ decays \cite{zref}, such as,  
$Z \rightarrow {e_i^\pm} e_j^\mp$ 
($\it{i,\,j} = e$, $\mu$, $\tau$ and hereafter, we will assume $i\neq j$ unless otherwise stated ), which have taken interest in view 
of future collider plans.
The present sensitivity of the measurement of branching ratio of 
$Z\rightarrow$ $e_i^\pm e_j^\mp$ 
at LEP is $\sim$ ${10}^{-5}$ whereas future linear colliders 
(NLC, JLC, Tesla GigaZ) will bring it down to $\sim$ ${10}^{-8}$ and 
thus the testability of such model will be increased due to 
higher sensitivity of measurement which
could be able to reveal new physics beyond the SM. 

In the present work, we calculate $\mu$AMM and LFV $Z$ decays 
in the Zee model keeping in view the other constraints arising due to 
$\mu$ decay and other LFV radiative lepton decays. 
We estimate $\mu$AMM and LFV $Z$ decays by utilizing the constraints 
on the parameter space obtained from the 
$\mu\rightarrow e{{\overline{\nu}}_e}\nu_\mu$ decay 
 and also from radiative $e_i\rightarrow e_j\gamma$ ( $\it{i}$ = $\mu$, $\tau$ ; $\it{j}$ =  $e, \mu$ ;  $\it{i\neq j}$ ) decays. 
Recent works in this path have been done \cite{ng} in which the Zee model 
have been investigated in view of recent  
$\mu$AMM experimental result and LFV decays.
In the present work, we restrict ourselves 
within the configuration of the minimal Zee model and we, particularly,
investigate different hierarchical cases of Zee $f_{ij}$ coupling subject to the present neutrino phenomenology. The plan of the paper is as follows: In Section II, 
we will first briefly review the Zee model, its basic 
interaction Lagrangian, the charged scalar 
mixing and the neutrino mass matrix. 
The constraints on the Zee $f_{ij}$  coupling due to 
$\mu\rightarrow e{{\overline{\nu}}_e}\nu_\mu$ , $e_i\rightarrow e_j\gamma$ 
decays and $\mu$AMM are discussed in Section III. 
The LFV $Z$ decays are calculated in Section IV and Section V 
contains summary of the present work.

\section{Brief Review of the Zee Model}
\label{sec:1}
\subsection{The interaction Lagrangian}

The interaction Lagrangian of the Zee model is given by 
$$
{\cal L}=\sum_{i,j}f_{ij}\overline{\ell}_{iL}i\tau_2
\ell^c_{jL}h^-
+c_{12}\phi_1^Ti\tau_2\phi_2h^-+
\sum_{i}y_i\overline{\ell}_{iL}\phi_2e_{iR}
+h.c.,
\eqno(2.1)
$$
where we have dropped the quark interaction terms, and 
lepton number conserving Higgs potential  terms. The lepton 
doublets are denoted as 
$\ell_{iL}$ ($i = 1,2,3$) 
with the definition  
$\ell^c_{iL}=(\ell_{iL})^c=C\overline{\ell}_{iL}^T $ 
 and  $\phi_a$ ($a$ = 1, 2) are the Higgs doublets and $h^-$ is
 a charged singlet scalar field.
The $f_{ij}$ terms can explicitly be written as follows:
\end{multicols}
\hspace{-0.5cm}
\rule{8.7cm}{0.1mm}\rule{0.1mm}{2mm}
\widetext
$$
\sum_{i,j}f_{ij}\overline{\ell}_{iL}i\tau_2\ell_{jL}^ch^-+h.c.
= \sum_{i,j}f_{ij}(\overline{\nu}_{iL}e_{jL}^c-
\overline{e}_{iL}\nu_{jL}^c)h^-+h.c. 
$$
$$
 = 2[f_{e\mu}(\overline{\nu}_{eL}\mu_L^c
-\overline{e}_L\nu_{{\mu}L}^c)
+f_{\mu\tau}(\overline{\nu}_{{\mu}L}\tau_L^c
-\overline{\mu}_L\nu_{{\tau}L}^c) 
+f_{{\tau}e}(\overline{\nu}_{\tau L}e_L^c
-\overline{\tau}_L\nu_{eL}^c)]h^-
+h.c.\, . 
\eqno(2.2)
$$
\hspace{9.1cm}
\rule{-2mm}{0.1mm}\rule{8.7cm}{0.1mm}
\begin{multicols}{2}
\narrowtext
\noindent
Since $\overline{e}_{iL}\nu_{jL}^c=\overline{\nu}_{jL}e_{iL}^c$, 
we can also 
re-express (2.2) as 
\end{multicols}
\hspace{-0.5cm}
\rule{8.7cm}{0.1mm}\rule{0.1mm}{2mm}
\widetext
$$
2[f_{e\mu}(\overline{\mu}_L\nu_{eL}^c-\overline{e}_L\nu_{{\mu}L}^c)
+f_{\mu\tau}(\overline{\tau_L}\nu_{{\mu}L}^c-\overline{\mu}_L\nu_{{\tau}L}^c)
+f_{{\tau}e}(\overline{e}_L\nu_{{\tau}L}^c-\overline{\tau}_L
\nu_{eL}^c)]h^-+h.c.
\ .
\eqno(2.3)
$$
\hspace{9.1cm}
\rule{-2mm}{0.1mm}\rule{8.7cm}{0.1mm}
\begin{multicols}{2}
\narrowtext
\noindent
The $c_{12}$ term can explicitly be expressed as follows:
$$
c_{12}\phi_1^Ti\tau_2\phi_2h^-+h.c.
=c_{12}(\phi^+_1\phi^0_2-\phi_1^0\phi_2^+)h^-+h.c.
\eqno(2.4)
$$
We define
$$
\phi^0_a={\frac{1}{\sqrt{2}}}(v_a+H_a^0-i\chi_a^0)\ ,
\eqno(2.5)
$$

$$
\chi^0=\chi^0_1\cos\beta-\chi^0_2\sin\beta \ , \ \ \ 
\widetilde{\chi}^0=\chi_1^0\sin\beta+\chi_2^0\cos\beta \ ,
\eqno(2.6)
$$
$$
H^0=H_1^0\cos\beta-H^0_2\sin\beta\ , \ \ \ 
\widetilde{H}^0=H_1^0\sin\beta+H_2^0\cos\beta\ ,
\eqno(2.7)
$$
$$
\phi^+=\phi_1^+\cos\beta-\phi_2^+\sin\beta\ , \ \ \ 
\widetilde{\phi}^+=\phi^+_1\sin\beta+\phi_2^+\cos\beta\ ,
\eqno(2.8)
$$
$$
\tan\beta=v_1/v_2\ ,
\eqno(2.9)
$$
where $v_a$ are the vacuum expectation values (VEV's) of $\phi_a^0$, 
$\langle\phi^0_a\rangle=v_a/{\sqrt{2}}$.

By using the expressions (2.6)-(2.9), the $c_{12}$ term is expressed 
as follows:

$$
c_{12}(\phi_1^+\phi_2^0-\phi_2^+\phi_1^0)h^-+h.c.
$$
$$
={\frac{1}{\sqrt{2}}}c_{12}[\phi^+
(\widetilde{v}+\widetilde{H}^0-i\widetilde{\chi}^0) - 
\widetilde{\phi}^+(v+H^0-i\chi^0)]h^-+h.c.,
\eqno(2.10)
$$
where
$$
v=v_1\cos\beta-v_2\sin\beta=0,
\eqno(2.11)
$$
$$
\widetilde{v}=v_1\sin\beta+v_2\cos\beta={\sqrt{v_1^2+v_2^2}}
$$
$$
= \frac{2 M_W}{g} ={\sqrt{2}}\times174{\rm GeV}.
\eqno(2.12)
$$
The components $\widetilde{\phi}^+$ and $\widetilde{\chi}^0$ are absorbed 
into the gauge bosons $W^+$ and $Z^0$, so that they are not physical 
particles.  Therefore, the physical part in the $c_{12}$ term is only 
$$
{\frac{1}{\sqrt{2}}}c_{12}\widetilde{v}\phi^+h^-+ 
{\frac{1}{\sqrt{2}}}c_{12}\phi^+\widetilde{H}^0h^-+
h.c..
\eqno(2.13)
$$
The first term of Eq.(2.13) is a quadratic interaction term which 
induces the $\phi^+$-$h^+$ mixing as we show in 
the later part of this section. The second term is not relevant for our 
analysis.

The third term in the Lagrangian (2.1) is the  
conventional Yukawa coupling term\ :
$$
\sum_{i}y_{i}(\overline{\nu}_{iL}\ \overline{e}_{iL})
\left(\begin{array}{c}
\phi^+_2 \\
\phi^0_2
\end{array}\right)
e_{iR}
=\sum_{i}y_i(\overline{\nu}_{iL}\phi^+_2+\overline{e}_{iL}\phi^0_2)e_{iR}
$$
$$
={\frac{1}{\sqrt{2}}}\widetilde{v}\cos\beta\sum_{i}
y_i(\overline{e}_{iL}e_{iR})
-\sin\beta\sum_{i}y_i(\overline{\nu}_{iL}e_{iR})\phi^+
$$
$$
-{\frac{1}{\sqrt{2}}}\sum_{i}y_i(\overline{e}_{iL}e_{iR})
[\sin\beta(H^0-i\chi^0)-\cos\beta\widetilde{H}^0]
$$
$$
+({\rm unphysical}\ {\rm terms}),
\eqno(2.14)
$$
so that the charged lepton masses $m_{ei}$ are given by 
$$
m_{ei}={\frac{1}{\sqrt{2}}}y_i{\widetilde{v}}\cos\beta .
\eqno(2.15)
$$
The term $\sum_iy_i\sin\beta(\overline{\nu}_{iL}e_{iR})\phi^+$ together 
with the terms given in Eqs.(2.3) and (2.13) will contribute to the neutrino mass generation. 

\subsection{The charged scalar boson mixing}

The mass matrix for $(h^+,\phi^+)$ is given by 
$$
M^2=
\left(\begin{array}{cc}
M_h^2 & c_{12}{{\widetilde{v}}/{\sqrt{2}}} \\
c_{12}{{\widetilde{v}}/{\sqrt{2}}} & M_\phi^2
\end{array}\right)
\eqno(2.16)
$$
where $M_h^2$ and $M_\phi^2$ are the coefficients of the $h^+h^-$, 
$\phi^+\phi^-$ terms of the scalar potential.  
We define an orthogonal transformation between the $h^+$ and 
$\phi^+$ scalar fields as 
$$
\left(\begin{array}{c}
h^+ \\
\phi^+
\end{array}\right)
=
\left(\begin{array}{cc}
\cos\theta & \sin\theta \\
-\sin\theta & \cos\theta
\end{array}\right)
\left(\begin{array}{c}
H_1^+ \\
H_2^+
\end{array}\right),
\eqno(2.17)
$$
by which  we obtain
$$
\left(\begin{array}{cc}
h^+ & \phi^+
\end{array}\right)
\left(\begin{array}{cc}
M_h^2 & c_{12}{{\widetilde{v}}/{\sqrt{2}}} \\
c_{12}{{\widetilde{v}}/{\sqrt{2}}} & M_\phi^2
\end{array}\right)
\left(\begin{array}{c}
h^- \\
\phi^-
\end{array}\right)
$$
$$
=
\left(\begin{array}{cc}
H^+_1 & H_2^+ 
\end{array}\right)
\left(\begin{array}{cc}
M_1^2 & M_{12} \\
M_{12} & M_2^2
\end{array}\right)
\left(\begin{array}{c}
H_1^- \\
H_2^-
\end{array}\right),
\eqno(2.18)
$$
$$
M_1^2=
{\frac{M_h^2+M_\phi^2}{2}}+{\frac{M_h^2-M_\phi^2}{2}}\cos2\theta-
c_{12}{\frac{\widetilde{v}}{\sqrt{2}}}\sin2\theta,
\eqno(2.19)
$$
$$
M_2^2=
{\frac{M_h^2+M_\phi^2}{2}}-{\frac{M_h^2-M_\phi^2}{2}}\cos2\theta+
c_{12}{\frac{\widetilde{v}}{\sqrt{2}}}\sin2\theta,
\eqno(2.20)
$$
$$
M_{12}=
{\frac{M_h^2-M_\phi^2}{2}}\sin2\theta+c_{12}{\frac{\widetilde{v}}{\sqrt{2}}}
\cos2\theta.
\eqno(2.21)
$$
The diagonalization condition gives 
$$
\tan2\theta=
{\frac{2c_{12}\widetilde{v}/{\sqrt{2}}}{M_\phi^2-M_h^2}}= 
{\frac{-2{\sqrt{2}}c_{12}m_W/g}
{\sqrt{(M_1^2-M_2^2)^2-(2{\sqrt{2}}c_{12}m_W/g)^2}}}.
\eqno(2.22)
$$

\subsection{The neutrino mass matrix}

The neutrino mass is generated in the Zee model due to the charged scalar
exchange at the one loop level through explicit lepton number violation. 
The Zee neutrino mass matrix $M_\nu$ is given by the form as
$$
M_\nu =\pmatrix{0& \rho & \sigma \cr
               \rho & 0 & \lambda \cr
               \sigma & \lambda & 0}
\eqno(2.23)
$$
where $\rho  = {(M_\nu)}_{e\mu}$,  $\sigma = {(M_\nu)}_{e\tau}$, 
 $\lambda = {(M_\nu)}_{\mu\tau}$ and 
$$
{(M_\nu)}_{ij} = m_0 f_{ij} ({m_{e_j}}^2 - {m_{e_i}}^2)/m_\tau^2
\eqno(2.24)
$$
$$
m_0 = \frac{\sin 2\theta\, \tan\beta\, m_\tau^2}{32 \pi^2 
{\widetilde{v}}/\sqrt{2}} \ln\frac{M_1^2}{M_2^2}
\eqno(2.25)
$$
and $\sin2\theta$ and $\tan\beta$ are determined from Eqs.(2.22) 
and (2.9), respectively. 

It has been shown in Ref.[6] that the model 
can accommodate the atmospheric neutrino experimental result as well
 as candidature of neutrino as a hot dark matter component through 
the choice of $f_{ij}$ coupling as 
 $f_{e\mu}\sim f_{\mu\tau}>f_{e\tau}$ and the neutrino mixing matrix 
obtained as
$$
U \simeq \pmatrix{1&0&0\cr
             0& \frac{1}{\sqrt{2}}& -\frac{1}{\sqrt{2}}\cr
             0& \frac{1}{\sqrt{2}}& \frac{1}{\sqrt{2}}
}
\eqno(2.26)
$$
and such mixing matrix cannot accommodate solar neutrino 
experimental results. It has been advocated to add a 
sterile neutrino in the model to explain the solar neutrino 
experimental results \cite{zeeref}. 

However, another interesting option to 
explain both the solar and atmospheric neutrino experimental 
results, namely, bi-maximal neutrino mixing can arise in the 
Zee model [7] due to the choice of $f_{ij}$ coupling as  
$f_{e\mu}\gg f_{e\tau}\gg f_{\mu\tau}$ which 
leads to the mixing matrix as 
$$
U \simeq \pmatrix{\frac{1}{\sqrt{2}}&-\frac{1}{\sqrt{2}}&0\cr
             \frac{1}{2}& \frac{1}{2}& -\frac{1}{\sqrt{2}}\cr
             \frac{1}{2}& \frac{1}{2}& \frac{1}{\sqrt{2}}
}
\eqno(2.27)
$$
and the mass matrix resemblance to the form presented in Ref.[15]. 
An explicit realization of the above mentioned 
hierarchy of Yukawa coupling has been 
demonstrated in Ref.[9] due to the inclusion of a badly broken 
$SU(3)_H$ horizontal symmetry  
through a simple ansatz on the symmetry breaking effects. 
The $SU(3)_H$ symmetry breaking has been considered to be 
proportional to the transition matrix elements of the mass-matrix 
and thereby obtained the three Yukawa couplings  
$$
f_{e\mu} = [m_\tau/(m_\mu+ m_e)]f, $$
$$
f_{e\tau} = -[m_\mu/(m_\tau+ m_e)]f,\eqno(2.28)$$
$$
f_{\mu\tau} = [m_e/(m_\tau+ m_\mu)]f,
$$
which necessarily leads to the hierarchy required to explain 
the solar and atmospheric neutrino experimental results 
through bi-maximal mixing. The solar and the atmospheric 
neutrino experimental results are connected through the relation 
$$
\Delta m^2_{sol}/\Delta m^2_{atm}\simeq \sqrt{2}m_e/m_\mu 
= 6.7\times {10}^{-3}
\eqno(2.29)
$$
which is in excellent agreement with the experimental results.
However, in the present work, we will show that bi-maximal mixing 
hierarchy confronts with the hierarchy required to explain the excess 
value of $\mu$AMM in the Zee model.

\section{
Constraint from $\mu  \rightarrow \lowercase{e}
\overline{\nu}_{\lowercase{e}} \nu_\mu$ and 
$\lowercase{e}_{\lowercase{i}} \rightarrow \lowercase{e}_{\lowercase{j}} 
\gamma$ Decays
}
\label{sec:2}

\subsection{$\mu{\rightarrow}e\overline{\nu}_e \nu_\mu$ decay}

{}From the interactions (2.1), we obtain the effective four Fermi 
interaction
\end{multicols}
\hspace{-0.5cm}
\rule{8.7cm}{0.1mm}\rule{0.1mm}{2mm}
\widetext
$$
{\cal L}_{eff}=
{\frac{1}{\overline{M}^2}}[-2f_{e\mu}(\overline{e}_L\nu_{{\mu}L}^c)
+2f_{{\tau}e}(\overline{e}_L\nu_{{\tau}L}^c)]
[2f_{e\mu}(\overline{\nu}_{eL}^c\mu_L)-
2f_{\mu\tau}(\overline{\nu}_{{\tau}L}^c\mu_L)]+\cdots
$$
$$
={\frac{4}{\overline{M}^2}}[-f_{e\mu}^2(\overline{e}_L\nu_{{\mu}L}^c)
(\overline{\nu}_{eL}^c\mu_L)-f_{{\tau}e}f_{\mu\tau}
(\overline{e}_L\nu_{{\tau}L})(\overline{\nu}_{{\tau}L}^c\mu_L)
$$
$$
+f_{{\tau}e}f_{e\mu}(\overline{e}_L\nu_{{\tau}L})
(\overline{\nu}_{eL}^c\mu_L)+
f_{e\mu}f_{\mu\tau}(\overline{e}_L\nu_{{\mu}L}^c)
(\overline{\nu}_{{\tau}L}^c\mu_L)]
+\cdots,
\eqno(3.1)
$$
\hspace{9.1cm}
\rule{-2mm}{0.1mm}\rule{8.7cm}{0.1mm}
\begin{multicols}{2}
\narrowtext
\noindent
where
$$
{\frac{1}{\overline{M}^2}}=
{\frac{\cos^2\theta}{M_1^2}}+{\frac{\sin^2\theta}{M_2^2}} \ .
\eqno(3.2)
$$
By using the Fierz transformation
and the formula
$$
\overline{\psi}_{1L}^cO\psi_{2L}^c=\overline{\psi}_{2L}
(C^{-1}OC)^T\psi_{1L},
\eqno(3.3)
$$
we can rewrite the first term of Eq.(3.1) as
$$
{\cal L}_{eff}
={\frac{1}{2}}{\frac{4}{\overline{M}^2}}f_{e\mu}^2
(\overline{e}_L\gamma_{\mu}\nu_{eL})
(\overline{\nu}_{\mu L}\gamma^{\mu}\mu_L).
\eqno(3.4)
$$
\noindent
On the other hand, the conventional interaction for the 
$\mu{\rightarrow}e{{\overline{\nu}}_e}\,\nu_\mu$ decay is 
given by
$$
{\cal L}=
{\frac{G_F}{\sqrt{2}}}(\overline{e}\gamma^{\mu}(1-\gamma_5)\nu_e)
(\overline{\nu}_{\mu}\gamma^{\mu}(1-\gamma_5)\mu) +h.c. \ ,
\eqno(3.5)
$$
$$
{\frac{G_F}{\sqrt{2}}}=
{\frac{g^2}{8M_W^2}}=
{\frac{1}{4(\widetilde{v}/{\sqrt{2}})^2}} \ ,
\eqno(3.6)
$$
so that the relative ratio of the contribution from (3.4) to that from 
(3.6) is given by 
$$
\zeta={\frac{4f_{e\mu}^2/{8\overline{M}^2}}{
{G_F}/\sqrt{2}}}= 2f_{e\mu}^2 \left(
\frac{\widetilde{v}/\sqrt{2}}{\overline{M}}\right)^2 \ .
\eqno(3.7)
$$
The four-Fermi coupling in the Zee model comes out as 
$(G_F/\sqrt{2})(1 +\zeta)$ where the parameter $\zeta$ 
is determined from the deviation between the observed ``$G_F$" value from 
$\mu\rightarrow e{{\overline{\nu}}_e}\nu_\mu$ and  that from 
hadronic weak decays.
Smirnov and Tanimoto \cite{zeeref} have put the constraint 
$\zeta < 10^{-3}$
(so that the model does not destroy the agreement in the 
electroweak precision tests), and they have obtained
$$
\xi f_{e\mu}^2 < 1.37 \times 10^{-4}, 
\eqno(3.8)
$$
where 
$$
\xi = \left( \frac{M_Z}{\overline{M}}\right)^2 \, .
\eqno(3.9)
$$ 
Furthermore, the second, third and the fourth terms of 
the effective Lagrangian give rise to the other possible 
$\mu$ decay modes as 
$\mu\rightarrow {{\overline{\nu}}_\tau}e\nu_\tau$,
 $\mu\rightarrow {{\overline{\nu}}_e}e\nu_\tau$,
 $\mu\rightarrow {{\overline{\nu}}_\tau}e\nu_\mu$ respectively, 
which also give the same final state signal as  
$\mu\rightarrow e\,\, +\,\, {\rm ``{missing\,\, energy}}"$. 
In a similar way, the constraints obtained from those processes 
are estimated with $\zeta^2 < {10}^{-3}$ as  
$$
\xi f_{\mu e} f_{e \tau}, \,\,\, \xi f_{\mu \tau} f_{e \tau}, \,\,\, 
\xi f_{\mu e} f_{\mu \tau} < 4\times 10^{-3} \ .
\eqno(3.10)
$$ 

\subsection{$e_i \rightarrow e_j \gamma$ decay}

The non-standard radiative decay \cite{petcov} $e_i \rightarrow e_j \gamma$ 
arises in the Zee model due to  $\gamma$-emission from the 
Zee boson, $\gamma$-emission from the initial state lepton 
and $\gamma$-emission 
from the final state lepton. 
In general, the gauge invariant total amplitude of the above 
process is obtained as
\end{multicols}
\hspace{-0.5cm}
\rule{8.7cm}{0.1mm}\rule{0.1mm}{2mm}
\widetext
$$
A(e_i\rightarrow {e_j} \gamma)=
{\frac{e f_{ik}f_{kj}}{96\pi^2{\overline{M}^2}}}
\overline{u}_j (p_2)
\sigma_{\mu\nu}[(m_i + m_j) + (m_i - m_j)\gamma_5] u_i(p_1)\varepsilon^{\mu}
(q)q^{\nu}\ ,
\eqno(3.11)
$$
\hspace{9.1cm}
\rule{-2mm}{0.1mm}\rule{8.7cm}{0.1mm}
\begin{multicols}{2}
\narrowtext
\noindent
with $q = p_1 - p_2$ and the decay width as 
$$
\Gamma (e_i \rightarrow e_j \gamma) = \frac{\alpha}{48\pi} 
\frac{m_i^5}{192 \pi^3}\left( \frac{f_{ik}f_{kj}}{
\overline{M}^2} \right)^2 ,
\eqno(3.12)
$$
where we have considered $m_i^2\gg m_j^2$.
To estimate the values of $f_{ij}$ we consider the ratio of the branching 
ratios of the decays $e_i\rightarrow {e_j} \gamma$
and $e_i\rightarrow e_j {{\overline{\nu}}_j}\nu_i$ as 
$$
\frac{{B(e_i \rightarrow e_j \gamma)}_{{\rm{Zee}}}}{{B(e_i \rightarrow
e_j \overline{\nu}_j \nu_i)}_{SM}} =  \frac{\alpha}{6\pi} 
\left( \frac{f_{ik}f_{kj}}{
[\overline{M}/(\tilde{v}/\sqrt{2})]^2} \right)^2 \ .
\eqno(3.13)
$$
The 
present experimental status of the above ratio is given by \cite{pdg} 
$$
\frac{B(\mu \rightarrow e \gamma)}{B(\mu \rightarrow
e \overline{\nu}_e \nu_\mu)} < 1.2\times 10^{-11} ,
\eqno(3.14)
$$
$$
\frac{B(\tau \rightarrow e \gamma)}{B(\tau \rightarrow
e \overline{\nu}_e \nu_\tau)} <  \frac{ 2.7 \times
10^{-6} }{(17.83\pm 0.06)\times 10^{-2}} , 
\eqno(3.15)
$$
$$
\frac{B(\tau \rightarrow \mu \gamma)}{B(\tau \rightarrow
\mu \overline{\nu}_\mu \nu_\tau)} <  \frac{  1.1 \times
10^{-6} }{(17.37 \pm 0.07)\times 10^{-2}} .
\eqno(3.16)
$$
Therefore, we obtain the constraints on the coupling constants as
$$
\xi |f_{\mu\tau}f_{\tau e}|< 
4.67\times {10}^{-5},
$$
$$
\xi |f_{e\mu}f_{\mu\tau}|< 
5.24\times {10}^{-2},
$$
$$
\xi |f_{\mu e}f_{e \tau }|< 
3.39\times {10}^{-2}
.
\eqno(3.17)
$$
We summarized the results obtained in Eqs.(3.8), (3.10) and (3.17) in Table I.

\noindent
\subsection{Anomalous muon magnetic moment ($\mu$AMM)}
\vskip .1in
\noindent
The excess value of $\mu$AMM recently reported by the E821 collaboration 
based on the theoretical calculation presented in Ref.[18] indicates the signal of new physics beyond the SM.
(However, it has been argued in Ref.[19] that the discrepancy 
between the theoretical and experimental values could be removed if other estimation of hadronic contribution to the photon propagator is considered ). 
The excess value of 
$$
\delta a_\mu = (43\pm 16)\times {10}^{-10}
\eqno(3.18)
$$
could arise in the Zee model due to charged scalar exchange at 
the one loop level. 
The diagrams are essentially same as 
the $e_i\rightarrow e_j\gamma$ process 
by regarding $i = j = \mu$. 
The extra contribution to $\delta a_\mu$ arises in the Zee model 
due to charged scalar exchange at the one loop level and it is estimated as 

$$
\delta a_\mu = \frac{\sum_{i = e, \tau} 
f_{\mu i}^2 m_\mu^2}{24 \pi^2 M_Z^2} \xi 
\ . 
\eqno(3.19)
$$
Therefore, if we regard that the value of (3.19) comes from the extra contribution due to charged scalar exchange we obtain the constraints on the couplings as 
$$
\xi ( f_{\mu e}^2 + f_{\mu\tau}^2) = 7.67\times {10}^{-1}.
\eqno(3.20)
$$
where we have considered the central value of $\delta a_\mu$.
The previously obtained value of $\xi f_{\mu e}^2$ from 
$\mu\rightarrow {{\overline{\nu}}_e} e \nu_\mu$ decay given in Eq.(3.8) 
is too low to explain the large value of $\delta a_\mu$, whereas 
the higher value of $\xi f_{\mu\tau}^2$ $\sim$ $7.67\times {10}^{-1}$ 
is not in conflict with the previous results given in Eqs.(3.10) and (3.17). 
It is to be noted that the hierarchy of $f_{ij}$ coupling required 
to explain such an excess value of $\mu$AMM ($f_{\mu\tau}\gg f_{e\mu}$) 
confronts with the hierarchy required to explain bi-maximal neutrino 
mixing as discussed in Section II.C. Thus, the upshot of our analysis is that
if we consider seriously the experimental result of $\mu$AMM experiment 
we have to give up bi-maximal neutrino mixing pattern in the Zee model.

\section{LFV $Z$ Decay's}

In this section, we calculate LFV $Z$ decays, 
$Z\rightarrow e_i^\pm e_j^\mp$
($i, j = e, \mu, \tau$ and $i\neq j$) which have taken much interest in view of  future 
colliders \cite{zref}. The explicit LFV coupling of the Zee model gives 
rise to such processes which is an immediate consequence of the non-zero 
neutrino mass. The diagrams of LFV $Z$ decay process have been given 
in Fig.~1. The total contribution of all diagrams can be written as 
\end{multicols}
\hspace{-0.5cm}
\rule{8.7cm}{0.1mm}\rule{0.1mm}{2mm}
\widetext
$$
A(Z(p)\rightarrow \overline{e}_i(p_1)e_j(p_2))
= \frac{g f_{ik}f_{kj}}{2\cos\theta_W} K
\overline{u}_j(p_2)\gamma_{\mu}(1 - \gamma_5 ) v_i(p_1)
\varepsilon^{\mu}(p) \ ,
\eqno(4.1)
$$
\hspace{9.1cm}
\rule{-2mm}{0.1mm}\rule{8.7cm}{0.1mm}
\begin{multicols}{2}
\narrowtext
\noindent
with $p = p_1 + p_2$ and 
$$
K = \frac{1}{2(4\pi)^2} \left\{ 
-2\sin^2\theta_W[\cos^2\theta{\cdot}F(\xi_1)+\sin^2\theta{\cdot}F(\xi_2)]
 \right\}
$$
$$
\left. +\cos^2\theta{\cdot}G(\xi_1)+\sin^2\theta{\cdot}G(\xi_2)
 -\sin^2\theta\cos^2\theta\cdot H(\xi_1, \xi_2) \right\} \ .
\eqno(4.2) 
$$
$$
F(\xi_i) = \int_0^1 dx \int_0^1 dy \,\,\, x \ln[1 - \xi_i xy(1-y)] \ ,
\eqno(4.3)
$$
$$
G(\xi_i)={\frac{3}{4}}+\int^1_0dx\int^1_0dy \,\,\, x \ln
\left[1-x-\xi_ix^2y(1-y)\right]
$$
$$
-\int^1_0dx\int^1_0dy
{\frac{\xi_ix^3y(1-y)}{1-x-\xi_ix^2y(1-y)}} \ ,
\eqno(4.4)
$$
$$
H(\xi_1, \xi_2) = F(\xi_1, \xi_2) + F(\xi_2, \xi_1) - 
F(\xi_1) - F(\xi_2)\ ,
\eqno(4.5)
$$
$$
F(\xi_1,\xi_2)=\int^1_0{dx}\int^1_0 dy \,\,\, x \ln \left[
1-(1 - {\frac{\xi_1}{\xi_2}})y-\xi_1xy(1-y)\right] \ ,
\eqno(4.6)
$$
and 
$$ 
\xi_1 = \frac{M_Z^2}{M_1^2} \ , \ \ \ 
  \xi_2 = \frac{M_Z^2}{M_2^2} \ .
\eqno(4.7)
$$
Since the mixing angle $\theta$ can also be written as 
$$
\sin^2\theta = \frac{M_\phi^2 - M_2^2}{M_1^2 - M_2^2}
             = \frac{\xi_2/\xi_\phi - 1}{\xi_2/\xi_1 - 1}
\eqno(4.8)
$$
where 
$$
\xi_\phi = \frac{M_Z^2}{M_\phi^2} \ ,
\eqno(4.9)
$$
and $\xi_\phi$ is related to 
$$
\xi\cdot\xi_\phi = \xi_1\cdot\xi_2 \ ,
\eqno(4.10)
$$
the factor $K$ is given as a function of the parameters 
$\xi_1$, $\xi_2$ and 
$\xi$. The approximate expressions of $F(\xi_i)$, $G(\xi_i)$ 
and  $H(\xi_1, \xi_2)$  are given in Appendix A. 
For $M_1^2> {\overline{M}}^2\gg M_\phi^2> M_2^2$, a dominant term in Eq.(4.2) 
is the $H(\xi_1, \xi_2)$ term  and we get 
\end{multicols}
\hspace{-0.5cm}
\rule{8.7cm}{0.1mm}\rule{0.1mm}{2mm}
\widetext
$$
K\simeq \frac{-1}{2{(4\pi)}^2}\sin^2\theta\cos^2\theta H(\xi_1, \xi_2)
$$
$$
= \frac{-1}{2{(4\pi)}^2}\frac{\xi_1\xi_2}{{(\xi_2 - \xi_1)}^2}
\left(\frac{\xi_2}{\xi_\phi} - 1 \right)
\left(1 - \frac{\xi_1}{\xi_\phi}\right)
\int_0^1 dx \int_0^1 dy \,\,\,x  \ln \left[ 1 + 
\frac{{(\xi_2 - \xi_1)}^2}{\xi_1 \xi_2}
\frac {y(1-y)}{[1- \xi_1 x y(1-y)][1 - \xi_2 x y (1-y)]} \right]
$$
$$
\simeq \frac{-1}{2{(4\pi)}^2}\frac{\xi_1}{\xi_2}
\left(\frac{\xi_2}{\xi_\phi} - 1 \right)
\left( \frac{1}{2}\ln\frac{\xi_2}{\xi_1} - 1 -\frac{1}{18}\xi_2\right)
\ .
\eqno(4.11)
$$
\hspace{9.1cm}
\rule{-2mm}{0.1mm}\rule{8.7cm}{0.1mm}
\begin{multicols}{2}
\narrowtext

The conventional lepton number conserving  $Z$ decay is described by
\end{multicols}
\hspace{-0.5cm}
\rule{8.7cm}{0.1mm}\rule{0.1mm}{2mm}
\widetext
$$
A(Z(p)\rightarrow \overline{e}_i(p_1)e_i(p_2))
= \frac{g}{2\cos\theta_W} 
\overline{u}_i(p_2)\gamma_{\mu}(g_V - g_A \gamma_5 ) v_i(p_1)
\varepsilon^{\mu}(p) \ , 
\eqno(4.12)
$$
\hspace{9.1cm}
\rule{-2mm}{0.1mm}\rule{8.7cm}{0.1mm}
\begin{multicols}{2}
\narrowtext
\noindent
$$
g_V= -\frac{1}{2}(1-4 \sin^2 \theta_W) \ , \ \ \ 
g_A =  -\frac{1}{2} \ ,
\eqno(4.13)
$$
$$
\Gamma(Z \rightarrow e_i^\pm e_i^\mp)= \frac{G_F}{\sqrt{2}}
\frac{M_Z^3}{6\pi} \left( g_V^2 +g_A^2 \right) \ .
\eqno(4.14)
$$
Therefore, the ratio of the branching ratios is given by
$$
R_{ij} = \frac{B(Z\rightarrow e_i^\pm e_j^\mp)}{
B(Z\rightarrow e_i^\pm e_i^\mp)} =
2 {\frac{{(\xi f_{ik}f_{jk})}^2}{g_V^2 + g_A^2}}
{\left({\frac{K}{{\xi}}}\right)}^2
\, .
\eqno(4.15)
$$
The experimental values of $B(Z\rightarrow e_i^\pm e_j^\mp)$ 
are as follows \cite{pdg} : 
$B(Z\rightarrow e^\pm\mu^\mp)< 1.7\times {10}^{-6}$, 
$B(Z\rightarrow e^\pm\tau^\mp)< 9.8\times {10}^{-6}$,
$B(Z\rightarrow \mu^\pm\tau^\mp)< 1.2\times {10}^{-5}$. 
Utilizing these experimental limits, we obtain 
$R_{e\mu}^{exp}< 5.05\times {10}^{-5}$,  
$R_{e\tau}^{exp}< 2.91\times {10}^{-4}$,  
$R_{\mu\tau}^{exp}< 3.56\times {10}^{-4}$ by using the branching ratio 
$B(Z\rightarrow \mu^\pm \mu^\mp) = 3.369\times {10}^{-2}$ \cite{pdg}.

Since our interest is in the maximum value of $R_{ij}$ under 
the constraints obtained in Eqs.(3.10) and (3.17), we have parameterized 
$R_{ij}$ in terms 
${(\xi f_{ik}f_{jk})}^2{\left({{K}/{\xi}}\right)}^2$ not by 
${(f_{ik}f_{jk})}^2K^2 $. 
As seen in Eq.(4.11), in the limit of $\xi\rightarrow 0$, 
the factor $K$ becomes vanishing while 
the value of $|K/\xi|$ increases as the value of $\xi_1$ 
decreases because $K/\xi$ is expressed as 
$$
{\frac{K}{\xi}}  
\simeq \frac{-1}{2{(4\pi)}^2}\frac{1}{\xi_2}
\left(1 - \kappa \right)
\left( \frac{1}{2}\ln\frac{\xi_2}{\xi_1} - 1 -\frac{1}{18}\xi_2\right) ,
\eqno(4.16)
$$
where 
$$
\kappa = \frac{ \xi_1}{\xi} = \frac{\xi_\phi}{\xi_2} 
= \frac{M_2^2}{M_\phi^2} \, .
\eqno(4.17)
$$
Therefore, for a small $\xi_1$ (but sizable $\xi_2$),
the factor $|K/\xi|$ is approximately proportional to 
$\ln(\xi_2/\xi_1)$ and $(1 - \kappa)$. 
Since $\xi_2 = \xi_\phi/\kappa$ ,  the value of $|K/\xi|$ has 
a maximum $|K/\xi|_{max}$ at $\kappa = \kappa_0 \simeq 1/2$ for a large 
$\xi_2/\xi_1$. 
We illustrate the behavior of $|K/\xi|$ versus $\kappa$ for three different 
values of $M_{\phi}$, $M_{\phi} = 200$ GeV, 500 GeV and 1 TeV 
(corresponding $\xi_{\phi} = $0.21, 0.033 and 0.0083 , respectively) in Fig.2.  Furthermore, for $\xi>10^{-2}$, the value of $\kappa_0$ deviates 
from $\kappa_0\simeq0.5$.  
This is due to the fact that the approximate expression (4.16) is valid 
only for $\xi\ll10^{-2}$.  
For $\xi<10^{-2}$, we can find that the expression (4.16) is 
in agreement with the direct numerical estimate within the deviation of 5 $\%$.

The value of $|K/\xi|$ is also highly dependent on the value of $\xi$. 
Since we want to obtain a value of $|K/\xi|$ as large as possible, 
we consider the value of $\xi$ as small as possible. We will estimate 
the lower bound of $\xi$ by using the perturbative unitarity bound on 
$f_{ij}$ coupling as $f_{ij}^2/4\pi < 1$ along with the experimental 
constraints on 
$\sqrt{\xi} f_{ij}$ given in Table II. It is to be noted that the experimental 
constraints on $\sqrt{\xi} f_{ij}$ depend on the assumptions for 
the hierarchical structures of $f_{ij}$. In Table II, we have considered 
the following three typical cases: 
(A) $f_{e\mu}^2$ = $f_{\mu\tau}^2$ = $f_{\tau e}^2$, 
(B) $f_{e\mu}^2$ $\gg$ $f_{\tau e}^2$ $\gg$ $f_{\mu\tau}^2$, and
(C) $f_{\mu\tau}^2$ $\gg$ $f_{e\mu}^2$ $\gg$ $f_{\tau e}^2$. 
The case (B) is motivated by the simultaneous explanation of 
the solar and atmospheric neutrino data as we discussed in Section II.C. 
The case (C) is motivated by the explanation of 
the excess value of $\mu$AMM.

In order to obtain a value of $|K/\xi|_{max}$ at $\kappa = \kappa_0$ as large 
as  possible, we want to take a value of $M_{\phi}$ as large as possible.  
However, it is unlikely that the value of $M_{\phi}$ is far from the 
electroweak scale.  For numerical estimate of $|K/\xi|_{max}$, we will take 
$M_{\phi}=500$ GeV.

(A) $f_{e\mu}^2$ = $f_{\mu\tau}^2$ = $f_{\tau e}^2$:  
In this situation, the most stringent bound comes from 
$\mu\rightarrow e\gamma$ decay as 
$$
{\sqrt{\xi}}|f_{e\mu}| = {\sqrt{\xi}}|f_{\mu\tau}| = {\sqrt{\xi}}|f_{\tau e}|
<0.68\times {10}^{-2}.
\eqno(4.18)
$$
We have listed those bounds on the ${\sqrt{\xi}} f_{ij}$ term for all the three cases in Table II.
In order to obtain a value of $R_{ij}$ as large as possible, 
we take the maximal value as 
${\sqrt{\xi}}|f_{ij}|\sim6.8\times10^{-3}$.  For such a value of 
${\sqrt{\xi}}|f_{ij}|$, we take a minimum value of $\xi$, 
$\xi\sim3.7\times10^{-6}$, under the perturbative unitary bound 
$f_{ij}^2/4\pi<1$.  Therefore, for a typical value $M_{\phi}=500$ GeV
($\xi_{\phi}$=0.033), together with $\kappa = \kappa_0 =$ 0.45, we 
estimate $\xi_1\sim1.6\times10^{-6}$ and $\xi_2\sim7.3\times10^{-2}$, 
which give the masses of the two charged scalars as $M_1\simeq 70$ TeV
 and $M_2\simeq 335$ GeV, respectively.  As seen in Fig.~2, the choice 
$\kappa = \kappa_0 = 0.45$ gives $|K/\xi|_{max}=1.11\times10^{-1}$, 
so that we predict
$$
R_{e\mu} = 
R_{e\tau} =
R_{\mu\tau}\simeq  2.1\times {10}^{-10}.
\eqno(4.19)
$$
The values of Eq.(4.19) are too small to observe even in the near future colliders. 
Moreover, the case does not give any interesting neutrino phenomenology, 
because the neutrino mixing in this case comes out as
$$
U \simeq \pmatrix{\frac{1}{\sqrt{2}}& \frac{1}{2}& \frac{1}{2}\cr
                  -\frac{1}{\sqrt{2}}& \frac{1}{2}& \frac{1}{2}\cr
                  0& -\frac{1}{\sqrt{2}}&  \frac{1}{\sqrt{2}}}
\eqno(4.20)
$$
with $m_{\nu_1}\ll |m_{\nu_2}| \simeq |m_{\nu_3}|$. 

(B) $f_{e\mu}^2$ $\gg$ $f_{\tau e}^2$ $\gg$ $f_{\mu\tau}^2$: 
The case is most interesting to us because it gives rise 
to bi-maximal mixing pattern as shown in Eq.(2.27) and in this case, 
the most serious bound on the $f_{ij}$ coupling is given in Eq.(3.8). 
The relation among the $f_{ij}$ coupling is given in Eq.(2.28) 
(which is necessary to explain reasonable value of 
$\Delta m^2_{solar}/\Delta m^2_{atm}$), so that it leads 
to the bounds as 
$$
{\sqrt{\xi}}|f_{e\mu}|< 1.17\times {10}^{-2}, \,\, 
$$
$$
{\sqrt{\xi}}|f_{e\tau}|< 4.08\times {10}^{-5}, \,\, 
$$
$$
{\sqrt{\xi}}|f_{\mu\tau}|< 2.84\times {10}^{-7} ,
\eqno(4.21)
$$
which predict  
$$
R_{e\mu}< 5.2\times {10}^{-24},
$$
$$
R_{e\tau}< 4.1\times {10}^{-19},
$$
$$R_{\mu\tau}< 1.8\times {10}^{-14}  ,
\eqno(4.22)
$$
where we have again utilized the constraints given in Eq.(4.21) giving 
$\xi\simeq1.1\times10^{-5}$, $|K/\xi|_{max}=9.73\times10^{-2}$ of 
$\kappa = \kappa_0 = 0.44$ and $\xi\simeq4.84\times10^{-6}$
($M_1\simeq41$ TeV), $\xi_2\simeq7.5\times10^{-2}$ ($M_2\simeq331$ GeV).   
Unfortunately, similar to the case (A), the values of $R_{ij}$  given in 
Eq.(4.22) are far away from the probing region of the future linear colliders. 

(C) $f_{\mu\tau}^2$ $\gg$ $f_{e\mu}^2$ $\gg$ $f_{\tau e}^2$: 
In order to  
give an explanation of the excess value of 
 $\mu$AMM, we take ${\sqrt{\xi}}|f_{\mu\tau}| = 0.876$ from Eq.(3.20) and 
 ${\sqrt{\xi}}|f_{\tau e}| \simeq 5.3\times {10}^{-5}$ from the Eq.(3.17) 
and ${\sqrt{\xi}}|f_{e\mu}| = 4.5 \times {10}^{-3}$ as obtained 
from Eq.(3.10). 
Utilizing the perturbative unitarity bound $f_{\mu\tau}^2< 4\pi$ 
we obtain the value of $\xi$ as $\xi = 6.1\times {10}^{-2}$.
For a typical choice of model parameters, $M_{\phi}= 500$ GeV, we obtain 
$|K/\xi|_{max}\simeq9.46\times10^{-3}$ at $\kappa = \kappa_0 = 0.11$, 
and we also get $\xi_1\simeq 6.71\times {10}^{-3}$ ($M_2\simeq1.1$ TeV),
$\xi_2\simeq 3.02\times {10}^{-1}$ ($M_2\simeq166$ GeV) .  
Therefore, we expect the maximal value of 
$$
R_{e\mu}< 1.6\times {10}^{-12},
$$
$$
R_{e\tau}< 1.1\times {10}^{-8},
$$
$$
R_{\mu\tau}< 4.2\times {10}^{-16}.  
\eqno(4.23)
$$
Although we have obtained a large value only for $R_{e\tau}$, the value 
$R_{e\tau}\sim 1.1\times10^{-8}$ is still two order away from the 
observable region of 
the future linear colliders.  Besides, since
 the case (C) gives the neutrino mixing given 
in Eq.(2.26) with the hierarchy of neutrino mass as 
$|m_{\nu_1}|\ll |m_{\nu_2}|\simeq |m_{\nu_3}|$, we must give up 
the explanation of the solar neutrino data within the framework 
of the three sequential neutrinos ($\nu_e$, $\nu_\mu$, $\nu_\tau$) 
in the Zee model. 

We summarized the main results of this section given in Eqs.(4.19), (4.22) 
and (4.23) in Table III.  

\section{Summary}
\vskip .1in
\noindent
In summary, we calculate LFV $Z$ decays  
in the context of the Zee model and see their observability in the future 
linear colliders subject to the existing bounds obtained from the $\mu$ decay, 
radiative $\mu$ and $\tau$ LFV decays and $\mu$AMM experimental result. 
We first constrain the parameter space 
by considering the bounds obtained from the $\mu$ decay, so that 
the total contribution to the four Fermi coupling $G_F$ should not 
exceed the measured experimental limits. We further have considered 
the bounds on the Yukawa couplings from the radiative charged lepton 
decays and $\mu$AMM. 
Both these constraints are given in Table I. 
The upshot of our analysis is that the hierarchy of coupling 
$f_{\mu\tau}^2\gg f_{e\mu}^2\gg f_{\tau e}^2$
needed to explain the excess value of anomalous muon magnetic moment is 
in conflict with the hierarchical pattern 
$f_{e\mu}^2 \gg f_{\tau e}^2 \gg f_{\mu\tau}^2$ which is required to 
obtain bi-maximal 
neutrino mixing in the Zee model with three active neutrinos in order to 
explain the solar and atmospheric neutrino experimental results.   
We investigate LFV  $Z$ decays in the context of three different choices 
of hierarchical relations of the Zee $f_{ij}$ coupling which are relevant 
to the present neutrino phenomenology. We find that among all the three 
decay modes of $Z$, only $Z\rightarrow e\tau$ decay gives rise to largest 
contribution to the ratio $R_{e\tau}$ $\sim$ ${10}^{-8}$ which is two order 
less than the accessible value to be reached by the future linear colliders 
only for the hierarchy of the $f_{ij}$ coupling $f_{\mu\tau}^2$ $\gg$ 
$f_{e\mu}^2$ $\gg$ $f_{\tau e}^2$ as addressed in the case (C) 
which cannot reconcile  the excess value of $\mu$AMM and the solar 
neutrino experimental result. Other possible LFV $Z$ decay modes for 
all the three cases are significantly small 
to be observed in the next linear colliders.

Although the values of the predicted branching ratios 
$B(Z{\rightarrow}e^{\pm}_ie^{\mp}_j$) are too small, the models (B) and (C) 
still remain as a 
promising candidate in the Zee model scenarios.  The case (B) can give 
the simultaneous 
explanation of the solar and atmospheric neutrino data, i.e., 
the bi-maximal mixing (2.26) together with 
${\Delta}m^2_{solar}/{\Delta}m^2_{atm}\simeq
{\sqrt{2}}m_e/m_{\mu}=6.7\times10^{-3}$, (2.29), although it cannot give an 
explanation of the observed excess of $\mu$AMM.  The excess may be understood 
by the contributions from SUSY partners.  On the other hand, the case (C) 
can give the simultaneous explanation of the atmospheric neutrino data and 
the observed excess of $\mu$AMM, although it cannot give an explanation of 
the solar neutrino data.  
The solar neutrino data may be understood \cite{zeeref} by an 
extended scenario with a sterile neutrino \cite{ma}, and so on.  
In the case (C), the neutrino mass matrix 
$M_{\nu}$ is given by the form 
$$
M_{\nu}=M_{\mu\tau}
\left(\begin{array}{ccc}
0 & \varepsilon_1 & \varepsilon_2 \\
\varepsilon_1 & 0 & 1 \\
\varepsilon_2 & 1 & 0 
\end{array}\right),
\eqno(5.1)
$$
which leads to 
$$
m_1\simeq-2\varepsilon_1\varepsilon_2M_{\mu\tau},
$$
$$
m_2\simeq-(1-\varepsilon_1\varepsilon_2)M_{\mu\tau},
$$
$$
m_3\simeq(1+\varepsilon_1\varepsilon_2)M_{\mu\tau},
\eqno(5.2)
$$
and
$$
U\simeq
\left(\begin{array}{ccc}
1 & {\frac{\varepsilon_2-\varepsilon_1}{\sqrt{2}}} & 
{\frac{\varepsilon_2+\varepsilon_1}{\sqrt{2}}} \\
-\varepsilon_2 & {\frac{1}{\sqrt{2}}} & {\frac{1}{\sqrt{2}}} \\
-\varepsilon_1 & -{\frac{1}{\sqrt{2}}} & {\frac{1}{\sqrt{2}}} 
\end{array}\right),
\eqno(5.3)
$$
where
$$
\varepsilon_1\simeq
{\frac{f_{e\mu}}{f_{\mu\tau}}}
\left({\frac{m_{\mu}}{m_{\tau}}}\right)^2 \ , \ \ \ 
\varepsilon_2\simeq
{\frac{f_{e\tau}}{f_{\mu\tau}}} \ .
\eqno(5.4)
$$
If we give up to obtain a large value of $B(Z{\rightarrow}e\tau$), 
the parameter $\kappa$ becomes free from the constraint $\kappa=\kappa_0$.  
Then, since $M_{\mu\tau}$ is given by
$$
M_{\mu\tau}\simeq2
{\frac{\sqrt{1-\kappa}}{\sqrt{\xi_2}}}{\sqrt{\xi}}f_{\mu\tau}
{\frac{\tan{\beta} \,\, m_{\tau}^2}{32\pi^2 \widetilde{v}/{\sqrt{2}}}}\ln
{\frac{\xi_2}{\xi_1}} \ ,
\eqno(5.5)
$$
we can obtain an arbitrary small value of $|M_{\mu\tau}|$ by taking 
$\kappa\rightarrow1$ ($M_2\rightarrow{M}_{\phi}$), with keeping the value 
${\sqrt{\xi}}f_{\mu\tau}$=0.876.

In conclusion, searches for LFV $Z$ decay are not useful to confirm
the two interesting models, (B) with $f_{e\mu}^2 \gg f_{\tau e}^2 \gg
f_{\mu\tau}^2$ and (C) with $f_{\mu\tau}^2 \gg f_{e\mu}^2 \gg
f_{\tau e}^2$ in the Zee's scenarios even at the near future colliders, 
where the model (B) can give the bi-maximal neutrino mixing together 
with the successful relation
$\Delta m^2_{solar}/\Delta m^2_{atm} \simeq \sqrt{2} m_e/m_\mu$
(although it fails to explain the observed excess of $\mu$AMM) and
the model (C) can give the simultaneous explanation of the
atmospheric neutrino data and the excess of $\mu$AMM (although
it cannot give any explanation of the solar neutrino data within
the framework of the three-flavor active neutrinos 
$(\nu_e, \nu_\mu, \nu_\tau)$).
Rather, an observation of the charged scalar (Zee scalar) associated
with lepton flavor violation decay will be important in the near future
colliders. 


\acknowledgements{
We are thankful to S. Ichinose and G. Bhattacharyya for many helpful 
comments and discussions. One of the authors (A.G) is supported by 
the Japan Society for Promotion of Science (JSPS) Postdoctoral 
Fellowship for Foreign Researches in Japan through Grant No. P99222.}


\begin{center}
{\large\bf{APPENDIX A}}
\end{center}
\vskip .1in
\begin{center}
{\large\bf{Behavior of $F(\xi_i)$, $G(\xi_i)$ and $H(\xi_1, \xi_2)$}}
\end{center}
\vskip .1in
\noindent
The function $F(\xi_i)$ is given by
$$
F(\xi_i)=\int^1_0dx\int^1_0dy{\cdot}x{\ln}[1-\xi_i xy(1-y)]
$$
\end{multicols}
\hspace{-0.5cm}
\rule{8.7cm}{0.1mm}\rule{0.1mm}{2mm}
\widetext
$$
={\frac{1}{4\xi_i^2}}\left[(4-5\xi_i)\xi_i+16
\left(\tan^{-1}{\sqrt{\frac{\xi_i}{4-\xi_i}}}
\right)^2-4{\sqrt{4-\xi_i}}{\sqrt{\xi_i}}(2-\xi_i)
\tan^{-1}{\sqrt{\frac{\xi_i}{4-\xi_i}}}
\right]
\eqno(A.1)
$$
\hspace{9.1cm}
\rule{-2mm}{0.1mm}\rule{8.7cm}{0.1mm}
\begin{multicols}{2}
\narrowtext
\noindent
\vskip .1in
For $\xi_i\ll1$, we obtain
$$
F(\xi_i)
=-{\frac{\xi_i}{18}}\left(1+{\frac{3}{40}}\xi_i+
{\frac{3}{350}}\xi_i^2+\cdots\right).
\eqno(A.2)
$$

The function $G(\xi_i)$ is given by
\end{multicols}
\hspace{-0.5cm}
\rule{8.7cm}{0.1mm}\rule{0.1mm}{2mm}
\widetext
$$
G(\xi_i) = - \frac{1}{2} 
+ \int_{x_0}^1 dx \cdot \frac{\xi_i x^2-2(1-x)}{ 
\sqrt{\xi_i}\sqrt{\xi_ix^2-4(1-x)} }
{\ln}
\frac{1+\sqrt{ \frac{\xi_ix^2-4(1-x)}{\xi_ix^2} } }{
1-\sqrt{\frac{\xi_ix^2-4(1-x)}{\xi_ix^2} } }
$$
$$
+ 2 \int_0^{x_0} dx \cdot \frac{2(1-x) -\xi_i x^2}{
\sqrt{\xi_i} \sqrt{4-4x -\xi_ix^2} }
\tan^{-1} \sqrt{\frac{\xi_ix^2}{4(1-x) -\xi_ix^2}} \ ,
\eqno(A.3)
$$
\hspace{9.1cm}
\rule{-2mm}{0.1mm}\rule{8.7cm}{0.1mm}
\begin{multicols}{2}
\narrowtext
\noindent
where $x_0$ = $(2/\xi_i)$$({\sqrt{1 + \xi_i}} -1)$.
The approximate expression of $G(\xi_i)$ obtained as 
$$
G(\xi_i) = -2.0827 \xi_i + 3.0687\xi_i ^2 -2.9787 \xi_i^3 
+1.1553 \xi_i^4 \cdots \ .
\eqno(A.4)
$$

We can rewrite the function $H(\xi_1, \xi_2)$ explicitly as 
follows:
\end{multicols}
\hspace{-0.5cm}
\rule{8.7cm}{0.1mm}\rule{0.1mm}{2mm}
\widetext
$$
H(\xi_1,\xi_2)=\int^1_0{dx}\int^1_0dy{\cdot}x \left\{ 
{\ln}\left[1- \frac{(1- \xi_1/\xi_2)y}{1- \xi_1 xy(1-y)} \right] 
+ {\ln}\left[1- \frac{(1- \xi_2/\xi_1)y}{1- \xi_1xy(1-y)} \right] 
\right\} 
$$
$$
= \int^1_0{dx}\int^1_0dy{\cdot}x  
{\ln}\left[1+ \frac{(\xi_1- \xi_2)^2y(1-y)}{
[\xi_1- \xi_1\xi_2 xy(1-y)][\xi_2- \xi_1 \xi_2 xy(1-y)]} \right] 
$$
$$
= \int^1_0{dx}\int^1_0dy{\cdot}x  
{\ln}\left[1+ B \frac{y(1-y)}{
[1- \xi_1 xy(1-y)][1- \xi_2 xy(1-y)]} \right]  \ ,
\eqno(A.5)
$$
\hspace{9.1cm}
\rule{-2mm}{0.1mm}\rule{8.7cm}{0.1mm}
\begin{multicols}{2}
\narrowtext
\noindent
where
$$
B= \frac{(\xi_1- \xi_2)^2}{\xi_1 \xi_2} = \left( \sqrt{\frac{\xi_1}{\xi_2}}
-\sqrt{\frac{\xi_2}{\xi_1}} \right)^2 \ .
\eqno(A.6)
$$
For $\xi_1/\xi_2 < 10^{-2}$, since $B \simeq \xi_2/\xi_1 > 10^2$, so that
we can approximate $H(\xi_1,\xi_2)$ as
$$
H(\xi_1, \xi_2) \simeq \int_0^1 dx \int_0^1 dy \cdot x \, \ln[1+ B y(1-y)]
-F(\xi_1) -F(\xi_2) 
$$
$$
\simeq  -\frac{1}{2} \left( 2 +\sqrt{1+\frac{4}{B}} \ln\frac{
\sqrt{1+4/B}-1}{\sqrt{1+4/B}+1} \right) -\frac{1}{18}(\xi_1+ \xi_2) 
$$
$$
\simeq -1 +\frac{1}{2} \ln\frac{\xi_2}{\xi_1} - \frac{1}{18}(\xi_1+ \xi_2) \ .
\eqno(A.7)
$$
Note that the expression (A.7) is valid only for $\xi_1/\xi_2<10^{-2}$.


\end{multicols}
\newpage

\vspace{5mm}
\widetext
\begin{table}
\begin{center}
{\bf{Table I}}
\vskip .20in
\begin{tabular}{|c|c|c|c|c|}
& 
$\xi f_{e\mu}^2$ & $\xi f_{\mu\tau} f_{\tau e}$ & $\xi f_{e\mu} f_{\mu\tau}$ & $\xi f_{\mu e} f_{e\tau}$\\
\hline
$\mu\rightarrow e{{\overline{\nu}}_i}\nu_j$ & $ 1.37 \times {10}^{-4}$ & $ 4\times{10}^{-3}$& $ 4\times{10}^{-3}$ & $ 4\times{10}^{-3}$\\
\hline
$e_i\rightarrow e_j \gamma$ & - & $ 4.67\times {10}^{-5}$& $5.24\times {10}^{-2}$& $3.39\times {10}^{-2}$\\
\hline
\end{tabular}
\caption
{ Experimental upper bounds on $\xi f_{ik} f_{kj}$  ($i, j, k = e, \mu, \tau $) from $\mu\rightarrow e{{\overline{\nu}}_i} \nu_j$  and 
$e_i\rightarrow e_j \gamma$
decays as discussed in Section II.A and B.}
\end{center}
\vskip .2in
\begin{center}
{\bf{Table II}}
\vskip .20in
\begin{tabular}{|c|c|c|c|c|
}
& 
${\sqrt{\xi}} |f_{e\mu}|$ & ${\sqrt{\xi}} |f_{e\tau}|$ & ${\sqrt{\xi}} |f_{\mu\tau}|$ & $\xi_{min}$\\ 
\hline
Case (A) : $|f_{e\mu}|$ = $|f_{e\tau}|$ = $|f_{\mu\tau}|$& $0.68\times {10}^{-2}$ & $0.68\times {10}^{-2}$  & $0.68\times {10}^{-2}$ & $3.7\times10^{-6}$\\
\hline
Case (B) : $|f_{e\mu}|$ $\gg$ $|f_{e\tau}|$ $\gg$ $|f_{\mu\tau}|$& $1.17\times {10}^{-2}$ &  $4.08\times {10}^{-5}$ & $2.84\times {10}^{-7}$ & 1$.1\times10^{-5}$ \\
\hline
Case (C) : $|f_{\mu\tau}|$$\gg$ $|f_{e\mu}|$ $\gg$ $f_{e\tau}|$ & $4.57\times {10}^{-3}$ & 5.33 $\times {10}^{-5}$ & $0.876$ & $6.1\times10^{-2}$\\
\hline
\end{tabular}
\caption
{ Possible maximal values of $f_{ij}$ coupling under the constraints 
given in Table I and the result of $\mu$AMM experiment.  
Also, the minimum value of $\xi$ under the constraints $f_{ij}^2/4\pi<1$ 
is listed for each case.  
}
\end{center}
\vskip .2in
\begin{center}
{\bf{Table III}}
\vskip .20in
\begin{tabular}{|c|c|c|c|c|c|c|c|}
& 
$\kappa_0$ & $|K/\xi|_{max}$ & $M_2$ & $M_1$ & 
$R_{e\mu}$ & $R_{e\tau}$& 
$R_{\mu\tau}$\\ 
\hline
Experimental upper bound & 
 & & & & $ 5.05 \times {10}^{-5}$ & $ 2.91 \times {10}^{-4}$  
& $ 3.56 \times {10}^{-4}$\\
\hline
Case (A) & 
0.45 & $1.1\times10^{-1}$ & 335 GeV & 70 TeV & 
$2.1 \times {10}^{-10}$ & $2.1 \times {10}^{-10}$  
& $2.1 \times {10}^{-10}$\\ \hline
Case (B) & 
0.44 & $9.7\times10^{-2}$ & 331 GeV & 41 TeV & $5.2 \times {10}^{-24}$ & 
$4.1 \times {10}^{-19}$ & $1.8 \times {10}^{-14}$\\
\hline
Case (C) & 
0.11 & 9.5$\times {10}^{-3}$ & 166 GeV & 1.1 TeV 
& $1.6\times {10}^{-12}$ & $1.1\times {10}^{-8}$ & 
$4.2 \times {10}^{-16}$\\
\end{tabular}
\caption
{ 
Prediction of the maximal values of $R_{ij}$  for the three different 
cases mentioned in Table II with the choices of model parameters 
as discussed in the text. The parameter values $M_1$ and $M_2$ are 
also listed.  
These value are fixed from the value of $\kappa_0$ at which $|K/\xi|$ 
takes the maximum $|K/\xi|_{max}$.
}
\end{center}

\label{T-qn}

\vglue.1in


\end{table}

\newpage

\mediumtext
\begin{figure}
\begin{center}

\includegraphics[width=15cm]{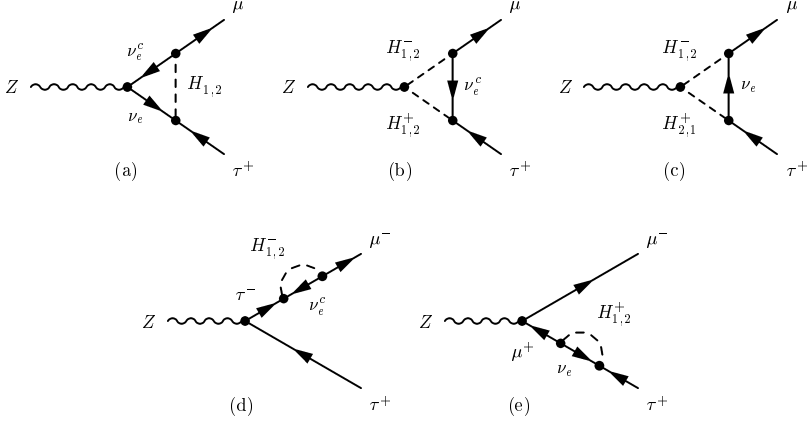}

\end{center}
\caption{
$Z\rightarrow \mu^-\tau^+$ decay in the Zee model due to charged 
scalar exchange. $Z\rightarrow \mu^-e^+$ and $Z\rightarrow e^-\tau^+$ 
decays also arise due to the same diagrams with appropriate 
replacement of the internal fermion.
}
\label{z-decay}
\end{figure}

\bigskip
\narrowtext
\begin{figure}
\begin{center}

\includegraphics[width=8.6cm]{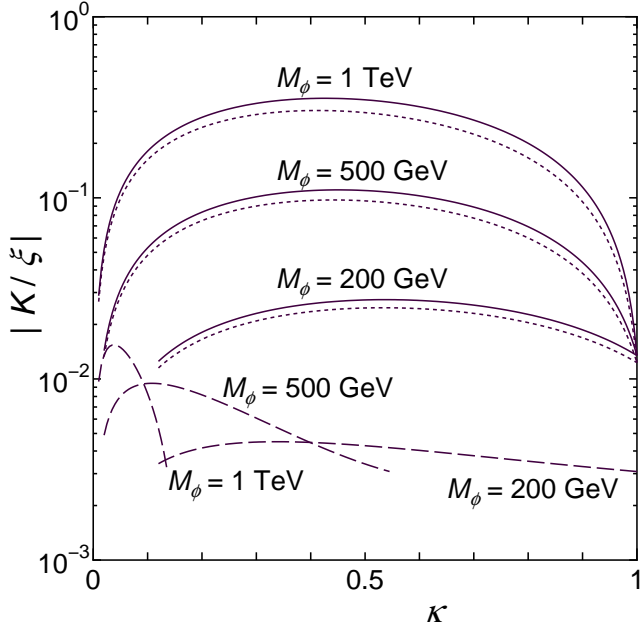}

\end{center}
\caption{
The factor $|K/\xi|$ is plotted against $\kappa$ for 
typical cases (A) (solid lines), (B) (dotted lines)  and (C) 
(dashed lines).
}
\label{k_xi}
\end{figure}
\end{document}